\documentclass[
 reprint,
 amsmath,
 pre,
 amssymb,
 nofootinbib,
 aps,
]{revtex4-2}

\usepackage{graphicx}
\usepackage{wasysym}
\usepackage{amsfonts}
\usepackage{ifsym}
\usepackage{braket}

\usepackage{amsmath,amssymb,amsfonts,amscd,mathrsfs,mathtools}
\usepackage{dsfont}

\usepackage{lipsum}
\usepackage{pifont}
\usepackage{float}
\usepackage{comment}

\usepackage{tikz-feynman}
 \tikzfeynmanset{compat=1.0.0}
\definecolor{darkblue}{rgb}{0.1,0.1,.7}

\makeatletter
\newlength{\apb@width}
\newcommand{\autoparbox}[2][c]{\settowidth{\apb@width}{#2}\parbox[#1]{\apb@width}{#2}}

\makeatother

\newcommand{\namedref}[2]{\hyperref[#2]{#1~\ref*{#2}}}

\newcommand{\mrm}[1]{\mathrm{#1}}

\newcommand{\be}{\begin{equation}}
\newcommand{\ee}{\end{equation}}

\newcommand{\reef}[1]{(\ref{#1})}

\def\bea#1\eea{\begin{align}#1\end{align}}

\newcommand{\eps}{\varepsilon}

\def\calO{{\cal O}}

\newcommand{\Csphere}{{}^\bullet\kern-1.2pt C}
\newcommand{\Ctorus}{{}^\circ\kern-1.2pt C}

\newcommand{\COMMENT}[1]{}

\newcommand{\neqa}{\nonumber\end{eqnarray}}

\newcommand{\<}{{\langle}}
\renewcommand{\>}{{\rangle}}

\newcommand{\re}{\relax{\rm I\kern-.18em R}}

\def\su2{{SU(2)}}

\def\eps{{\epsilon}}

\def\[{\left[}
\def\]{\right]}

\def\({\left(}
\def\){\right)}
\def\[{\left[}
\def\]{\right]}

\def\<{\langle}
\def\>{\rangle}

\def\i2{\frac{i}{2}}

\def\2F1{\,_2{\rm F}_1}

\usepackage{color}
\usepackage{graphicx}
\usepackage{dcolumn}
\usepackage{bm}
\usepackage{hyperref}

\usepackage{cancel}
\usepackage{ctable}
\usepackage{booktabs}
\newcolumntype{L}[1]{>{\raggedright\let\newline\\\arraybackslash\hspace{0pt}}m{#1}}
\newcolumntype{C}[1]{>{\centering\let\newline\\\arraybackslash\hspace{0pt}}m{#1}}
\newcolumntype{R}[1]{>{\raggedleft\let\newline\\\arraybackslash\hspace{0pt}}m{#1}}

\newcommand{\beq}{\begin{equation}}
\newcommand{\eeq}{\end{equation}}
\newcommand{\beqq}{\begin{equation*}}
\newcommand{\eeqq}{\end{equation*}}
\newcommand\beqa{\begin{eqnarray}}
\newcommand\eeqa{\end{eqnarray}}
\newcommand\beqaa{\begin{eqnarray*}}
\newcommand\eeqaa{\end{eqnarray*}}

\begin{document}

\title{
Reentrant $d=2+1$ Superfluidity at High Temperature}
\author{Subham Dutta Chowdhury}
\email{sdutta\_c@ictp.it}

\author{Joan Elias Mir\'o}
\email{eliasmiro@ictp.it}
 
\affiliation{\vspace{.2cm}    The Abdus Salam ICTP,    Strada Costiera 11, 34135, Trieste, Italy }

\begin{abstract}
We present a UV-complete model with a global $U(1)$ symmetry in $d=2+1$
dimensions that exhibits a
Berezinskii-Kosterlitz-Thouless (BKT) ordered phase at arbitrarily high temperatures.
A careful separation of scales yields a two-dimensional Euclidean effective field
theory whose construction remains perturbatively controlled even when its soft
modes become strongly coupled. We determine the BKT phase boundary by carefully
matching this theory to existing lattice Monte Carlo results.
The high-temperature phase supports gapless static modes and algebraically
decaying correlations even as $T\to\infty$.
\end{abstract}

\pacs{Valid PACS appear here} 

\maketitle


\section{Introduction}

Systems that undergo order--disorder phase transitions are usually
expected to be ordered at low temperature and disordered at high
temperature.
In Quantum Field Theory (QFT), this expectation is reflected in the typically
positive thermal corrections to scalar masses squared, which favor
symmetry restoration at high temperature.
However, this need not be the case.
Indeed, in ref.~\cite{Weinberg:1974hy}
S.~Weinberg  showed   that these corrections can be negative in theories with
several scalar fields. 
A symmetry can then remain broken at high temperature, or become broken
upon heating.

Recent constructions of local, UV-complete QFTs in $d=3+1$
realize symmetry nonrestoration at arbitrarily high
temperature~\cite{Chaudhuri:2020xxb,Chaudhuri:2021dsq,Bajc:2020gpa,Bajc:2026ppk,HarderEtAl2026Misalignment,Harder:2026jjr}.
High-temperature order has also been studied in lattice models through
entropic mechanisms~\cite{Han:2025eiw,zhang2025infinite,Hsin:2026ckp,Tsao:2026hle}.
In $2+1$ dimensions, persistent breaking of discrete symmetries has been
found in local models~\cite{Chai:2020zgq,Chai:2020onq,Liendo:2022bmv,Hawashin:2024dpp,Komargodski:2024zmt,Smolkin:2026wij,Chaudhuri:2026ges}.
Persistent breaking of continuous symmetries is also possible in nonlocal
QFTs with long-range interactions, including in $2+1$
dimensions~\cite{Chai:2021djc,Chai:2021tpt}.
Here we construct a local, UV-complete model in $2+1$ dimensions with
a small number of fields and quasi-long-range $U(1)$ order at arbitrarily
high temperatures.

Consider a complex scalar $\chi$ in $d=2+1$ spacetime dimensions with  negative mass squared
and weak
quartic self-interaction w.r.t. to the mass  parameter.
At zero temperature, its condensate spontaneously breaks
$U(1)$, producing a massless Goldstone mode.

At any nonzero temperature $T$, the long-wavelength static modes are
described by a two-dimensional Euclidean theory. The
Coleman--Mermin--Wagner theorem then forbids spontaneous breaking of the
continuous symmetry~\cite{Coleman:1973ci,Mermin:1966fe,Hohenberg:1967zz}.
Nevertheless, at sufficiently low temperature the weakly coupled model
lies deep in a Berezinskii--Kosterlitz--Thouless (BKT) phase.
The zero-temperature Goldstone field survives as its massless compact
phase field, supporting algebraic correlations and superfluidity, with
vortices bound in vortex--antivortex
pairs~\cite{Berezinskii:1970pzv,Berezinskii:1972fet,Kosterlitz:1973xp,Kosterlitz:1974nba}.
Thus an arbitrarily small temperature restores the symmetry while
preserving the massless  phase field.

Further heating eventually drives the ordinary scalar theory into a
disordered phase: positive thermal corrections to the mass parameter lead
to a gapped phase, with correlators that decay exponentially rather than
algebraically.

In this work we address how to realize a BKT phase at arbitrarily high $T$ in a 
local UV-complete model, with a small number of fields. 
To that scope,  we couple $\chi$
to a Conformal Field Theory (CFT) through a parametrically weak portal
${\cal O}_\Delta|\chi|^2$, where ${\cal O}_\Delta$ is a scalar operator
of  dimension $\Delta$.
The sign of the portal coupling is chosen so that the thermal one-point
function $\langle{\cal O}_\Delta\rangle_T\propto T^\Delta$ induces a
negative scalar mass shift at leading order.
 
The mechanism can be summarised as follows.    At  high temperatures, the negative conformal contribution to the scalar mass squared dominates the positive scalar thermal correction for $\Delta>1$. The resulting dimensionless phase stiffness grows as
$K(T)\sim(T/m)^{\Delta-1}$ so that algebraic correlations decay more slowly as the temperature is increased. We show that with all the microscopic couplings held fixed and sufficiently weak, this entire mechanism is controlled
in the window $1<\Delta<3/2$. The upper bound ensures both stability of the  vacuum effective potential and parametrically decreasing backreaction on the conformal sector at high temperature.  

The energy operator of the three-dimensional Ising CFT has
$\Delta_\epsilon\simeq1.4126$~\cite{Kos:2016ysd} and provides a concrete
 construction of high $T$ superfluidity  with only a  small number of fields,
which in turn admits a simple  lattice description.

 Our analysis proceeds by integrating out    the screened CFT sector and the nonzero scalar
Matsubara modes to obtain a local two-dimensional complex scalar theory.
Comparing its couplings with the nonperturbative BKT
boundary~\cite{Prokof_ev_2001} determines its phase.
For a sufficiently weak portal, the disordered regime survives between
the low- and high-temperature BKT phases, giving at least two BKT
transitions and the  reentrant sequence BKT--disordered--BKT.

\section{The model}

We consider the model
\be\label{coupledaction}
 S=S_{U(1)}+S_{\mrm{CFT}}
 +h\int d^3x\;m^{2-\Delta}{\cal O}_\Delta(x)|\chi(x)|^2 \, ,
\ee
where we denote by $S_{U(1)}$ the \emph{ordinary $U(1)$ model},
\be\label{ordinary_3d_action}
 S_{U(1)}=\int d^3x\bigg[
 |\partial_\mu\chi|^2
 +\frac g2\bigg(|\chi|^2-\frac{m^2}{g}\bigg)^2
 \bigg]\,.
\ee
Here $g$ and $m$ both have mass dimension one, $[g]=[m]=1$.
Throughout this work, we take the dimensionless coupling
$\lambda\equiv g/m\ll1$ to be small. 
The putative $d=2+1$ CFT described
by $S_{\mrm{CFT}}$ contains a scalar operator $O_\Delta$ of dimension
$\Delta>1$. We couple the two sectors
through the portal interaction with $h$.

We must ensure boundedness of the effective potential of $\chi$. 
After  integrating out the CFT, it is given by
\begin{align}\label{veff}
V_{\rm eff}
=
\frac g2\bigg(|\chi|^2-\frac{m^2}{g}\bigg)^2
+\mathcal E_{\rm CFT}(J)-\mathcal E_{\rm CFT}(0) \, .
\end{align}
to leading order in $g$. We have defined  $J\equiv  h m^{2- \Delta} |\chi|^2$ and $\mathcal E_{\rm CFT}(J){-}\mathcal E_{\rm CFT}(0)
=A_\pm|J|^{\frac{3}{3-\Delta}}$.
Here $\mathcal E_{\rm CFT}(J)$ denotes the vacuum-energy density,
defined from $-\log Z_{\rm CFT}(J)/\text{Vol}$,
  the constants $A_\pm$ correspond to the two
signs of $J$, and scaling dictates that $|J|$ dependence. 
At large $|\chi|$, the positive scalar quartic dominates the CFT
contribution  provided $\Delta <3 /2$. This ensures   
boundedness of the effective potential.

The deep-UV fixed point of \reef{coupledaction} is the product of a free
complex scalar and the CFT. 
The condensate  sources ${\cal O}_\Delta$, introducing the scale
$M_{\rm CFT}\sim m(|h|/\lambda)^{1/(3-\Delta)}\ll m$. If this relevant
deformation has a massive endpoint, $M_{\rm CFT}$ sets the corresponding
gap. 
In other words, depending on whether the source $v^2 O_{\Delta_\text{UV}}$
gaps or triggers a flow to another CFT, the leading irrelevant IR interaction for $k\ll M_\text{CFT}$ will be $(\partial \theta)^4$ or perhaps $(\partial \theta)^2 O_{\Delta_\text{IR}}$.~\footnote{When the leading irrelevant   interaction is $(\partial \theta)^2 O_{\Delta_\text{IR}}$, this is an example of an irrational soft limit for the goldstone mode, with a $2\rightarrow 2$ amplitude $M(s)= s^{\Delta_\text{IR}}+ \cdots$. See ref.~\cite{Brauner:2022ymm} for considerations on fractional soft limits.}
In the window
$M_{\rm CFT}\ll k\ll m$, the EFT is approximately a compact Goldstone
boson times 
a weakly deformed CFT,   with the two sectors coupled only through irrelevant  operators.

\medskip

We now turn to the theory at finite temperature $T$,
where the Euclidean theory lives on
$S^1_{1/T}\times\mathbb R^2$.
We assume that all degrees of freedom other than the scalar zero mode
have static screening gaps of order $2 \pi nT$ or larger. 
 Integrating them out
at spatial momenta
\be
k/(2\pi T) \ll1 \, . \label{onecond}
 \ee
 gives a local two-dimensional action. 
 The condition  \reef{onecond} will be assumed throughout.
 With the normalization
$\chi_{ n=0}=\sqrt{T}\,\varphi$,
\begin{equation}\label{fixed_2d_complex_action}
 S_2=\int d^2x\left[
 |\nabla\varphi|^2+r_2(\mu)|\varphi|^2
 +\frac{u_2}{2}|\varphi|^4+\delta{\cal L}_2
 \right].
\end{equation}
Here $\mu$ is the two-dimensional $\overline{\mathrm{MS}}$
subtraction scale, and $\delta{\cal L}_2$ contains   additional local interactions.
 The coefficients
$r_2$ and $u_2$ are obtained by matching to the $2+1$-dimensional theory;
their expansions depend on the temperature regime.
The matching equations are perturbative as long as hard  thermal loops are small, i.e. 
\be \frac{g}{2\pi T}\ll1 \quad \text{and} \quad  |h|\left(\frac{m}{2\pi T}\right)^{2-\Delta}\ll1 \label{second}
\ee
Furthermore, provided   these conditions   are satisfied, ${\cal \delta  L}_2$  in \reef{fixed_2d_complex_action} is negligible.

\medskip

At low temperature  but within the regime of (\ref{onecond}) and (\ref{second}), the expansion is  organised about the
radial minimum $|\varphi|^2\simeq m^2/(gT)$. In the window
$M_{\rm CFT}\ll T\ll m$,  the leading
dynamics is governed by the ordinary $U(1)$ theory.~\footnote{See below the analysis of CFT back-reaction.}
The  leading parametric behavior  of $r_2$ and $u_2$  take the form,
\begin{align}
 r_2(T;\mu)
 &=
 -m^2,
 \qquad
 u_2(T)
 =
 gT \, . 
 \label{ordinary_leading_matching_lowT}
\end{align}
The aforementioned corrections and the  portal
contributions are subleading in this analysis.

\medskip

For $T\gg m$ and the favorable sign $b_{\mathcal O}h<0$,
the leading matching gives
\begin{align}
 r_2(T;\mu)
 &=
 r_{2,U(1)}(T;\mu)
 -|b_{\mathcal O}h|m^{2-\Delta}T^\Delta\, ,  \label{cfttt} 
 \\
 r_{2,U(1)}(T;\mu)
 &=
 -m^2+\frac{gT}{\pi}\ln\frac{T}{\mu}
\ ,  \quad 
 u_2(T)
 =
 gT
 \label{ordinary_leading_matching}
\end{align}
where we suppress further higher cumulant and higher loop corrections.
At leading order in $h$, the CFT one-point function shifts only the
quadratic coefficient and drives it toward negative values at high
temperature.  
 
 \medskip

We address to which extent our approximation are valid.
 We analyse the regime in which the thermal mass is
dominated by  the CFT-induced term
$-|b_{\cal O}h|\,m^{2-\Delta}T^\Delta$.
In particular we  identify the conditions under which the
back-reaction of the scalar background on the thermal CFT state
remains parametrically small, so that higher cumulants are sub-leading
and the small-source expansion is controlled.

To determine the thermal background of $\chi$, we expand
the CFT contribution around zero source:
\bea
&\mathcal E^T_{\rm CFT}(J)-\mathcal E^T_{\rm CFT}(0) \\
=&J\langle\mathcal O\rangle_T
-\frac{J^2}{2}\int_0^{1/T} d\tau\int d^2\vec x\,
\langle\mathcal O(\tau,\vec x)\mathcal O(0)\rangle_{T,c}
+O(J^3) \, .  \nonumber
\eea
where $\mathcal E^T_{\rm CFT}$ denotes the thermal free-energy and  $\langle {\cal O}(x)\rangle_T=b_{ \cal O}T^\Delta$.   The second cumulant is ultraviolet finite for $\Delta<3/2$ and
infrared finite if the connected correlator decays exponentially
at large spatial separation \footnote{
Thermal compactification need not gap a generic CFT; the
three-dimensional Ising CFT considered below is nevertheless expected to
develop a thermal gap,  consistent with the analysis of Ref.~\cite{Iliesiu:2018zlz}.}. We assume a regular thermal
small-source expansion also for the higher cumulants.

For  $T \gg m$, the first cumulant correction to \eqref{veff} dominates the quadratic part of the effective potential for $\Delta>1$.
The second and higher cumulants can be ignored as the dimensionless coupling $|J(\chi_T)|/T^{3-\Delta}$ that suppresses the higher cumulants decreases for $\Delta<\frac{3}{2}$,
\begin{align}\label{backreaction}
|J(\chi_T)|/T^{3-\Delta} \simeq \frac{b_{\mathcal O} h^2 m}{g} \left( \frac{m}{T}\right)^{3- 2 \Delta} \to 0 \, ,
\end{align}
where $\chi_T$ denotes the thermal saddle and $\langle\mathcal O\rangle_T= b_{\mathcal O} T^\Delta$. At low $T$, such that $T/m\ll1$, higher cumulants can likewise be neglected
provided $|h|/\lambda\ll(T/m)^{3-\Delta}$. 

\medskip

Thus the same bound $\Delta<3/2$ ensures quartic domination of the
vacuum effective potential at large field, ultraviolet finiteness of the
second thermal cumulant, and parametrically decreasing CFT backreaction as
$T\to\infty$.

A concrete candidate for this mechanism is the $d=2+1$ Ising CFT.
Its leading $\mathbb Z_2$-even scalar, the energy operator $\epsilon$,
has scaling dimension $\Delta_\eps\simeq1.4126$~\cite{Kos:2016ysd},
within the required window $1<\Delta_\eps<3/2$, and a nonzero thermal
one-point function
$\langle\eps\rangle_T=b_\eps T^{\Delta_\eps}$, with $b_\eps=0.63(7)$~\cite{Iliesiu:2018zlz}.
Choosing ${\cal O}_\Delta=\eps$ and the portal sign so that
$h b_\eps<0$ then gives the negative thermal mass contribution
needed for the mechanism.

\section{The phases of the 2D  theory}
\label{sec:fixed_static_problem}

Here we determine whether the leading static theory
\reef{fixed_2d_complex_action} is gapped or exhibits quasi-long-range
order. Throughout this section, we study the family of static theories
parametrized by $u_2$ and $m_0^2\equiv r_2(\mu_0)$, holding these
parameters fixed in each theory. Here $\mu_0>0$ is a fixed reference mass,
$u_2>0$, and $m_0^2$ can have either sign. The quadratic coefficient runs as
$
r_2(\mu)= m_0^2+ \frac{u_2}{2\pi}\log(\mu_0^2/\mu^2)
$.
 The phase diagram can be formulated without specifying their
temperature dependence. Thermal matching will select a trajectory
through this phase diagram, as discussed in
Sec.~\ref{sec:thermal_trajectories}.

In two-dimensions $[\varphi]=0$ and $[r_2]=[u_2]=2$.  The theory is
superrenormalizable.  The field and
quartic coupling have no logarithmic ultraviolet renormalization, while the
quadratic coefficient is additively renormalized.  In the
$\overline{\mrm{MS}}$ convention used below, the mass parameter runs as
$d r_2(\mu)/d\log\mu^2=-u_2/(2\pi)$
(see Appendix~\ref{app:PRS_matching}).
Consequently $2r_2/u_2$ alone is not invariant under a change of
subtraction scale.  A convenient scale-independent coordinate on the
two-dimensional theory space is ${\cal R}$,  defined by
\be\label{R_definition}
 {\cal R}\equiv
 \frac{2r_2(\mu)}{u_2}
 +\frac1\pi\log\frac{2\mu^2}{u_2} \, . 
\ee
At $\mu^2=u_2/2$, this reduces to ${\cal R}=2r_2(\mu)/u_2$.
Equivalently,
$
 {\cal R}=\frac{2m_0^2}{u_2}
 +\frac1\pi\log\frac{2\mu_0^2}{u_2}$.

\medskip

\emph{Deep massive regime:}
The deep massive regime is characterized by ${\cal R}\gg1$.
Choosing the subtraction scale $\mu=\mu_0$, with $\mu_0^2$ of order
$m_0^2>0$, we have $r_2(\mu_0)=m_0^2$ and no large perturbative
logarithms. The gap is therefore
$
 M_{\mrm{gap}}^2=m_0^2
 \left[1+O({\cal R}^{-1})\right]$. 
 The two-point function
$\langle\varphi(x)\varphi^\dagger(0)\rangle$ decays exponentially as
$|x|\to\infty$, with correlation length $M_{\mrm{gap}}^{-1}$.  The finite
mass gap and exponential decay of correlations therefore place the theory
deep in the disordered regime.

\medskip

\emph{Deep algebraic regime and the phase EFT:}
The deep algebraic regime is characterized by ${\cal R}\ll-1$.
Choosing the subtraction scale $\mu=\mu_0$, with $\mu_0^2$ of order
$|m_0^2|$ and $m_0^2<0$, we have $r_2(\mu_0)=m_0^2$ and
$|m_0^2|/u_2\gg1$. The potential then has a well-developed Mexican-hat
minimum. The classical saddle and radial mass are, to leading order,
$v^2\simeq-{\cal R}$ and $m_\rho^2\simeq-u_2{\cal R}$.
The expansion in radial fluctuations is controlled by
$1/v^2\sim -1/{\cal R}\ll1$.

The compact phase of the complex field admits vortex defects.
Writing $\varphi=(v+\rho)e^{i\theta}/\sqrt2$ and integrating out the
radial mode at momenta $k\ll m_\rho$, we obtain the phase EFT
\be\label{phaseaction}
 S_\text{EFT}=\frac{K}{2}\int d^2x\big[
 (\partial\theta)^2
 -\frac{1}{m_\rho^2}(\partial\theta)^4+\ldots
\big] \, , 
\ee
where $ \theta\sim\theta+2\pi$.
Here $K$ is the stiffness matched at the radial scale, with
$K=v^2[1+\calO(v^{-2})]$.
Thus the spin-wave ($\mrm{sw}$) corrections are local derivative operators suppressed by
$m_\rho$.  Smooth configurations alone are not the complete EFT, because the
phase is compact.
An isolated vortex is a saddle in a fixed-winding sector; with
neutral boundary conditions, vortices instead enter through interacting
vortex--antivortex configurations weighted by their core actions.
The organization of the compact phase EFT as a neutral
vortex gas is standard \cite{Kosterlitz:1974nba,PhysRevB.16.1217}.  Let
$Z_{\mrm{sw}}=\int{\cal D}\theta\,e^{-S_{\mrm{sw}}}$.  With neutral boundary
conditions, the finite-volume fugacity expansion begins with a
vortex--antivortex pair and may be organized as
$
 Z=Z_{\mrm{sw}}\left[1+\calO(y_1^2)+\ldots\right]$, 
 where  $y_1=\calO\big(e^{-c_{\mrm{core}}v^2}\big)$, with $ c_{\mrm{core}}>0
$. 
The infrared theory therefore approaches the compact-boson fixed line, where
\be\label{fixed_algebraic_correlator}
 \left\langle e^{i\theta(x)}e^{-i\theta(0)}\right\rangle
 \sim |x|^{-\eta}\, ,
 \qquad \eta=\frac{1}{2\pi K}\, .
\ee
There is algebraic order (a.k.a.  \emph{quasi long range order}) but no nonzero infinite-volume order parameter.

\medskip

\emph{The critical point:} The determination of $K$ and $y_1$ from $(r_2,u_2)$ is semiclassically
controlled only in the deep algebraic regime.  Near the transition, radial
and vortex-core fluctuations are both order one.  
In other words, for ${\cal R}=\calO(1)$ the two-dimensional static theory is
strongly coupled; from the $2+1$-dimensional thermal perspective this is the
soft sector, and locating its critical point requires a nonperturbative
numerical treatment.

The asymptotic long-distance
flow is nevertheless BKT.  Writing $K_{\mrm R}$ for the renormalized infrared
stiffness, the critical trajectory satisfies $K_{\mrm R}=2/\pi$, with
$\Delta_v=\pi K_{\mrm R}=2$.
What cannot be obtained from the small-fugacity equations alone is the
location of that trajectory in the static $(r_2,u_2)$ parameter space.
Prokof'ev, Ruebenacker, and Svistunov (PRS) locate this trajectory by Monte
Carlo simulation of a regulated $|\psi|^4$ model and BKT finite-size scaling
\cite{Prokof_ev_2001}.  We translate the PRS conventions into the
$\overline{\mrm{MS}}$ convention of \eqref{fixed_2d_complex_action} and find
\be\label{PRS_main_result}
 {\cal R}_c=-0.821(10) \, . 
\ee
The detailed lattice-to-dimensional-regularization conversion is given in
Appendix~\ref{app:PRS_matching}.

Eq.~\eqref{PRS_main_result} 
is the nonperturbative position of the  complex scalar theory on the BKT
separatrix.
The phase assignment is therefore ${\cal R}<{\cal R}_c$ for the BKT phase and ${\cal R}>{\cal R}_c$ for the massive phase.
At the scale $\mu^2=u_2/2$, the negative value of ${\cal R}_c$
means that the transition occurs while $r_2(\mu)<0$. As
${\cal R}$ increases through ${\cal R}_c$, the unit-vortex operator changes
from irrelevant to relevant. The numerical value of
${\cal R}_c$ is fixed by the full nonperturbative radial and
vortex-core dynamics.

\section{From a static point  to \\ a thermal trajectory}
\label{sec:thermal_trajectories}

A $2+1$-dimensional UV completion supplies a one-parameter family of static
theories of the form \reef{fixed_2d_complex_action},
$
 \big\{r_2(T;\mu),u_2(T)\big\}
$.
The phase, either quasi-long-range order or disorder, is diagnosed by
${\cal R}(T)=2r_2(T;\mu)/u_2(T)
+\pi^{-1}\log[2\mu^2/u_2(T)]$, which is \reef{R_definition} evaluated along the one-parameter trajectory
supplied by the UV completion and parametrized by $T$.
The PRS critical surface therefore appears as the horizontal line
${\cal R}={\cal R}_c$ when ${\cal R}(T)$ is plotted against temperature. We
define $\delta(T)\equiv{\cal R}(T)-{\cal R}_c$, so that
\be\label{delta_phase_assignment}
 \delta(T)<0:\ \hbox{BKT} \, ,
 \qquad
 \delta(T)>0:\ \hbox{massive} \, ,
\ee
while the BKT transition occurs at
$\delta(T_{\mrm{BKT}})=0$.

For the ordinary $U(1)$ model, i.e. the theory \reef{ordinary_3d_action}, 
\be\label{ordinary_delta_endpoints}
 \delta_{U(1)}{=}
 \begin{cases}
 -\dfrac{2m^2}{gT}
 +O\bigl(\log[\tfrac{m^2}{gT}]\bigr)
 \rightarrow-\infty \, ,
 & T\to0,\\[7pt]
 \dfrac1\pi\log(\tfrac{T}{g})+O(1)
 \rightarrow+\infty \, ,
 & T\to\infty.
 \end{cases}
\ee 
Indeed, at low temperature eq.~\reef{ordinary_leading_matching_lowT} gives
${\cal R}_{U(1)}(T)
=-2m^2/(gT)+\calO[\log(m^2/(gT))]$.
At high temperature, hard matching in the $\overline{\mrm{MS}}$ scheme gives
\reef{ordinary_leading_matching}, therefore one obtains  
${\cal R}_{U(1)}(T)
=\pi^{-1}\log(T/g)+\calO(1)$.
By continuity, the trajectory therefore crosses the critical surface at
least once.

For the favorable portal sign $b_{\cal O} h<0$ and
$1<\Delta<3/2$, the CFT-coupled trajectory shares the ordinary
low-temperature endpoint but has the high-temperature asymptotics
\be\label{CFT_delta_highT}
 \delta_{U(1)+\mrm{CFT}}{=}-\frac{2|b_{\cal O} h|}{\lambda}
 \left(\frac{T}{m}\right)^{\Delta-1}
 +O\bigl(\log\tfrac{T}{g}\bigr)
 \rightarrow-\infty
\ee
as $T\to\infty$.

Indeed, eqs.~(\ref{cfttt},\ref{ordinary_leading_matching}) give
$\delta_{U(1)+\mrm{CFT}}(T)
=\delta_{U(1)}(T)
-(2|b_{\cal O } h|/\lambda)
(T/m)^{\Delta-1}+\ldots$.
Since $\Delta>1$, the portal contribution vanishes as $T\to0$, so
the low-temperature endpoint is unchanged.  
As argued above, the upper bound $\Delta<3/2$ ensures, through \eqref{backreaction}, that the CFT backreaction decreases in this limit and the leading matching remains controlled.
 
\begin{figure}[t]
 \centering
 \includegraphics[width=\columnwidth]{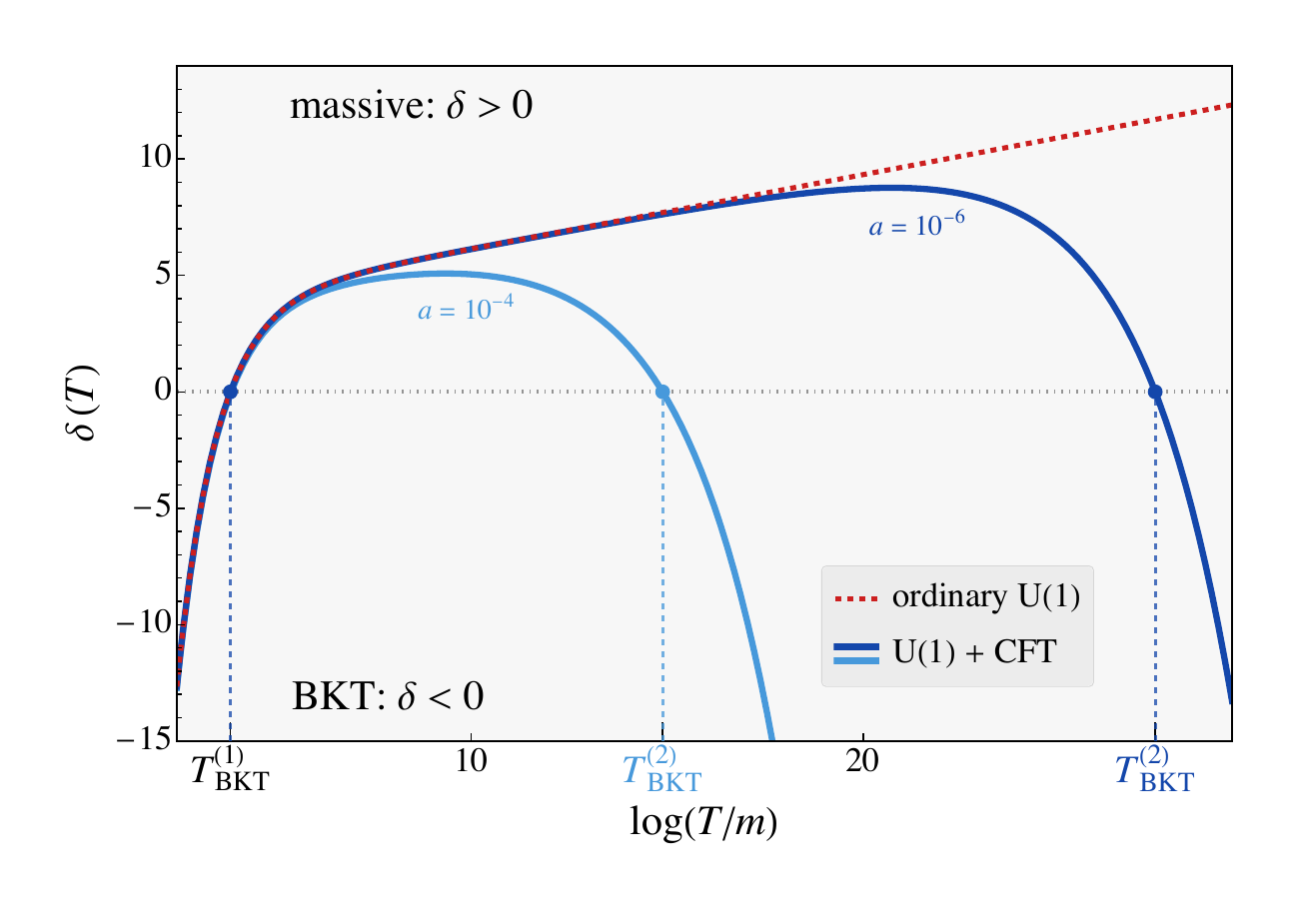}
 \caption[Representative ordinary and CFT-coupled thermal trajectories.]{We plot an analytic
 example,
 $\displaystyle \delta_{U(1)+\mrm{CFT}}^{\mrm{rep}}(t)
 ={-}\tfrac{2}{\lambda t}
{ }+\tfrac1\pi\log\!\left(1{+}\tfrac{t}{\lambda}\right)
{ }-\tfrac{2a}{\lambda}t^{\Delta-1}{+}c_0$, which reproduces  the
 asymptotic endpoint behaviours.    Here $c_0$ is
 a vertical offset.  For illustration,  $\lambda{=}0.01$,
 $a{=}10^{-4},10^{-6}$, $\Delta{=}1.4$, and $c_0{=}1.5$.}
 \label{fig:representative_thermal_trajectories}
\end{figure}

 Equation~\eqref{CFT_delta_highT} places the coupled trajectory in the BKT
region at  arbitrary high $T$.  For a sufficiently small but nonzero
favorable portal coupling, the intermediate massive region of the ordinary
trajectory persists, so that
$\max_T\delta_{U(1)+\mrm{CFT}}(T)>0$.
By continuity, the trajectory then crosses the critical surface at least
twice, yielding the reentrant phase sequence BKT-massive-BKT.  The three
deep regimes can be parametrically controlled, while the crossings
themselves are fixed nonperturbatively by the PRS critical value.

For a sufficiently strong  portal
contribution relative to the scalar self-interaction, with the favorable sign, the intermediate
massive interval can disappear, leaving a BKT phase at every nonzero
temperature. Note that the asymptotic high $T$ suppression of CFT backreaction follows from $\Delta<3/2$
and holds for any fixed couplings $h$ and $\lambda>0$, independently of
their relative size.

\medskip 

To illustrate the endpoint asymptotics, we plot representative trajectories
against $x=\log(T/m)$ in
Fig.~\ref{fig:representative_thermal_trajectories}.
We remark  that the couplings ensuring perturbative    hard matching conditions \reef{second}  and the local
expansion parameter \reef{onecond} remain small throughout the plotted
range, while the soft modes become strongly coupled near $\delta=0$.
In the plotted curves, $t=T/m$ and $a=|b_{\cal O}h|$;
setting $a=0$ gives the ordinary trajectory.
With $\delta={\cal R}-{\cal R}_c$, the $a=10^{-6}$ curve reaches
${\cal R}\simeq7.95$ at its maximum and ${\cal R}<-13$ at both ends of the
plotted range. The curves are schematic and do not determine the transition temperatures.

For sufficiently weak portal coupling, the coupled trajectory closely follows
the ordinary $U(1)$ asymptotics over a broad temperature range.
A smaller portal coupling extends this range to higher temperatures.
The reentrant BKT phase at high temperature is driven by the
asymptotic dominance of the CFT contribution.

\section{Discussion}

 We have established a local and UV-complete QFT, given by \reef{coupledaction}, that features   a reentrant BKT--disordered--BKT phase sequence
as the temperature increases.

A central feature of the mechanism is the simultaneous high-temperature limit
\bea
K(T)&\sim (T/m)^{\Delta-1}\longrightarrow\infty\, ,\\
|J(\chi_T)|/T^{3-\Delta}&\sim (T/m)^{2\Delta-3}\longrightarrow0\, .
\eea
For $1<\Delta<3/2$, increasing temperature enhances the phase stiffness $K$
while suppressing the backreaction on the thermal CFT state.

The $d=3$ Ising CFT provides a concrete example through its
energy operator, whose scaling dimension is $\Delta_\eps\simeq1.4$. Thus providing an example of a model with a small number of fields featuring BKT phase at arbitrarily high temperature.
 
Our model supports a BKT phase at arbitrarily high temperatures with only a few fields; the simplest realization involves just $\chi$ and a critical Ising field. It would therefore be particularly interesting to construct a simple microscopic lattice model exhibiting the same reentrant BKT behaviour.

A few directions deserve further investigation.
It would be interesting  to investigate the real-time dynamics of the high-temperature BKT phase. Dimensional analysis suggests that the damping of phase fluctuations at  long wavelength decreases with temperature. A careful computation of the hydrodynamic modes and their dissipation would allow us to test this expectation.

Finally, it would be interesting to analyse non-Abelian generalisations, in particular whether varying the temperature can drive transitions between WZW or their coset CFTs and massive phases. Gauged versions of these theories and of the Abelian model studied here would also be interesting.

\noindent\textbf{Note added.}
This work was first presented at the University of Geneva on
8 October 2026~\cite{EliasMiroGeneva2026}.
While this manuscript was being completed 
Ref.~\cite{Hawashin:2026smh} appeared. 
There it is shown that,
for sufficiently large $N$, the biconical fixed point of an
$O(N)\times U(1)$ model supports a BKT phase, and hence
superfluidity, at all nonzero temperatures.

\section*{Acknowledgements}

We are grateful to  Olivier Delouche for useful discussions and comments on the draft. 
We thank Brando Bellazzini, James Ingoldby,
Markus Luty, Francesco Riva,  Junchen Rong, Giovanni Villadoro and Yuan Xin for useful discussions. 
S.D.C. and J.E.M.  are  supported by the European Research Council, grant agreement n. 101039756. 

\addcontentsline{toc}{section}{References}
\bibliographystyle{utphys}
\bibliography{high_temperature_bkt_summary}

\clearpage
\appendix
\onecolumngrid
\makeatletter
\let\set@footnotewidth\set@footnotewidth@one
\let\compose@footnotes\compose@footnotes@one
\onecolumn@grid@setup
\count\footins=1000\relax
\makeatother

\section{Lattice and dimensional regularisation matching}
\label{app:PRS_matching}

This appendix derives the critical value quoted in
\eqref{PRS_main_result}.  PRS~\cite{Prokof_ev_2001} study a classical complex field on a square
lattice, with Hamiltonian
\be\label{app_PRS_unit_lattice_hamiltonian}
 H=\sum_{\mathbf k\in{\rm BZ}}
 [E(\mathbf k)-\nu_{\rm lat}]|\psi_{\mathbf k}|^2
 +\frac U2\sum_i|\psi_i|^4,
 \qquad
 E(\mathbf k)=4-2\cos k_x-2\cos k_y,
\ee
and statistical weight $e^{-H}$. Here we adopt their simulation units
$T=a=1$, $M=1/2$.

They locate the transition by varying $\nu_{\rm lat}$ until the measured
superfluid density satisfies the Nelson--Kosterlitz criterion
$
 n_s(L)=\frac1\pi
$. 
In these units, the continuum phase stiffness is $K_{\rm R}=2n_s$,\footnote{
For the phase-gradient energy expressed in terms of the superfluid density,
see for instance the review Ref.~\cite{HadzibabicDalibard2011}.
} so this
criterion corresponds to the universal critical value $K_{{\rm R},c}=2/\pi$.
The total density and chemical potential at this point define $n_c(L)$ and
$\nu_c(L)$; Repeating this procedure at several lattice sizes, they extract the infinite-volume critical density and chemical potential using the finite-size dependence predicted by BKT theory.
In the continuum limit of their lattice model, PRS obtain the lattice
critical chemical potential
\be\label{app_PRS_lattice_result}
 \nu_{\mrm{lat},c}=\frac{U}{2\pi}
 \log\left(\frac{4A_\mu}{U}\right),
 \qquad A_\mu=211(6).
\ee

Introduce a physical lattice spacing $a$ and define the continuum complex
field by $\Phi(x_i)=\psi_i$, with $x_i=a\,i$. No field rescaling is required
in the PRS simulation units. Since $E(\mathbf k)=\mathbf k^2+O(k^4)$,
the continuum action and the PRS critical line take the form
\be\label{app_PRS_QFT_bare_critical_r}
 S=\int d^2x\left[|\nabla\Phi|^2+r_0|\Phi|^2
 +\frac{u_2}{2}|\Phi|^4\right] \quad ,  \quad\quad
 \frac{2r_{0,c}}{u_2}
 =-\frac1\pi\log\left(\frac{4A_\mu}{u_2a^2}\right).
\ee
Here $A_\mu=211(6)$ is the constant quoted in
\eqref{app_PRS_lattice_result}, and the bare parameters are
$r_0=-\nu_{\rm lat}/a^2$ and $u_2=U/a^2$, in the normalization used in
the main text.
To relate the bare mass $r_0$ to the renormalized parameter $r_2(\mu)$,
we compare the tadpole contribution in lattice and dimensional
regularization.

The lattice and dimensionally regularized tadpoles are
\bea
 I_{\mrm{lat}}(s)
 &=\int_{\mrm{BZ}}\frac{d^2p}{(2\pi)^2}
 \frac1{\widehat p^{\,2}+s}
 =\frac1{4\pi}\log\frac{32}{a^2s}
 +\calO\!\left(a^2s\log(a^2s)\right),
 \label{app_lattice_tadpole}\\
 I_{\mrm{DR}}(s)
 &=\left(\frac{e^{\gamma_E}\mu^2}{4\pi}\right)^\epsilon
 \int\frac{d^{2-2\epsilon}p}{(2\pi)^{2-2\epsilon}}
 \frac1{p^2+s}
 =\frac1{4\pi}\left[\frac1\epsilon+\log\frac{\mu^2}{s}\right]
 +\calO(\epsilon).
 \label{app_DR_tadpole}
\eea
Here $s>0$ is a common auxiliary mass squared regulating the infrared,
$\widehat p^{\,2}=E(a\mathbf p)/a^2$ is the lattice kinetic kernel,
and ${\rm BZ}=[-\pi/a,\pi/a]^2$.
The lattice result is expanded for $a^2s\to0$.
In dimensional regularization, $d=2-2\epsilon$ and $\mu$ is the
$\overline{\mrm{MS}}$ subtraction scale; after subtraction, the finite
part is $I_{\mrm{DR}}^{\mrm{ren}}(s)=(4\pi)^{-1}\log(\mu^2/s)$.
With the normalization in \eqref{app_PRS_QFT_bare_critical_r}, the tadpole
self-energy is $2u_2 I$. Matching the quadratic terms therefore gives,
in the continuum limit,
\be\label{app_mass_conversion}
 r_2(\mu)
 =r_0+2u_2\left[I_{\mrm{lat}}(s)-I_{\mrm{DR}}^{\mrm{ren}}(s)\right]
 =r_0+\frac{u_2}{2\pi}\log\frac{32}{a^2\mu^2}.
\ee
The auxiliary mass $s$ cancels. Combining the PRS critical line
\eqref{app_PRS_QFT_bare_critical_r} with the definition
\eqref{R_definition} then yields
\be\label{app_Rc_from_lattice}
 {\cal R}_c
 =\frac{2r_{0,c}}{u_2}
 +\frac1\pi\log\frac{64}{u_2a^2}
 =\frac1\pi\log\frac{16}{A_\mu}
 =-0.821(10).
\ee
The cutoff $a$ and subtraction scale $\mu$ cancel from ${\cal R}_c$.
In PRS notation, $\xi_\mu\equiv A_\mu/16=13.2(4)$, so the same result
reads ${\cal R}_c=-\log\xi_\mu/\pi$.

\end{document}